\documentclass[aps,prl,10pt,twocolumn,superscriptaddress,showpacs,showkeys]{revtex4-1} 

\usepackage{graphicx}
\usepackage{hyperref}
\usepackage{setspace}
\usepackage{amsmath}
\usepackage{siunitx}
\renewcommand{\paragraph}[1]{\emph{#1}.---}

\begin{document}

\title{Pinning-induced folding-unfolding asymmetry in adhesive creases}

\author{Michiel A.J. van Limbeek}
\affiliation{Max Planck Institute for Dynamics and Self-Organization, 37077 G\"ottingen, Germany}
\author{Martin H. Essink}
\affiliation{Physics of Fluids Group, Mesa+ Institute, University of Twente, 7500 AE Enschede, The Netherlands}
\author{Anupam Pandey}
\affiliation{Biological and Environmental Engineering Department, Cornell University, Ithaca, NY 14853, USA}
\author{Jacco H. Snoeijer}
\email{j.h.snoeijer@utwente.nl}
\affiliation{Physics of Fluids Group, Mesa+ Institute, University of Twente, 7500 AE Enschede, The Netherlands}
\author{Stefan Karpitschka}
\email{stefan.karpitschka@ds.mpg.de}
\affiliation{Max Planck Institute for Dynamics and Self-Organization, 37077 G\"ottingen, Germany}

\date{\today}

\begin{abstract}
The compression of soft elastic matter and biological tissue can lead to creasing, an instability where a surface folds sharply into periodic self-contacts. Intriguingly, the unfolding of the surface upon releasing the strain is usually not perfect: small scars remain that serve as nuclei for creases during repeated compressions. Here we present creasing experiments with sticky polymer surfaces, using confocal microscopy, which resolve the contact line region where folding and unfolding occurs. It is found that surface tension induces a second fold, at the edge of the self-contact, which leads to a singular elastic stress and self-similar crease morphologies. However, these profiles exhibit an intrinsic folding-unfolding asymmetry that is caused by contact line pinning, in a way that resembles wetting of liquids on imperfect solids. Contact line pinning is therefore a key element of creasing: it inhibits complete unfolding and gives soft surfaces a folding memory.
\end{abstract}

\maketitle

Creases are ubiquitous to nature and can readily be observed by closing ones' hand or bending ones' arm: Soft tissue responds to compression by folding into deep valleys of self-contacting skin~\cite{gent1999rubber,hong2009formation,dervaux2012review}. The morphology of mammalian brains~\cite{tallinen2014gyrification,mota2015cortical,tallinen2016cortical,karzbrun2018organoids} or tumors~\cite{dervaux2011tumors} is dominated by creases that emerge from tissue growth under constraint conditions.
Similarly, polymer coatings in technological applications may suffer from creasing due to swelling~\cite{onuki1989theory,tanaka1994swelling,trujillo2008hydrogels,kim2009patterns,benamar2010swelling},
but also compressed elastomers~\cite{gent1999rubber,ghatak2007kink,mora2011instability,cai2012elastomer} and viscoelastic liquids~\cite{mora2011instability} display this instability.

\begin{figure}[tb!]%
	\centering\includegraphics{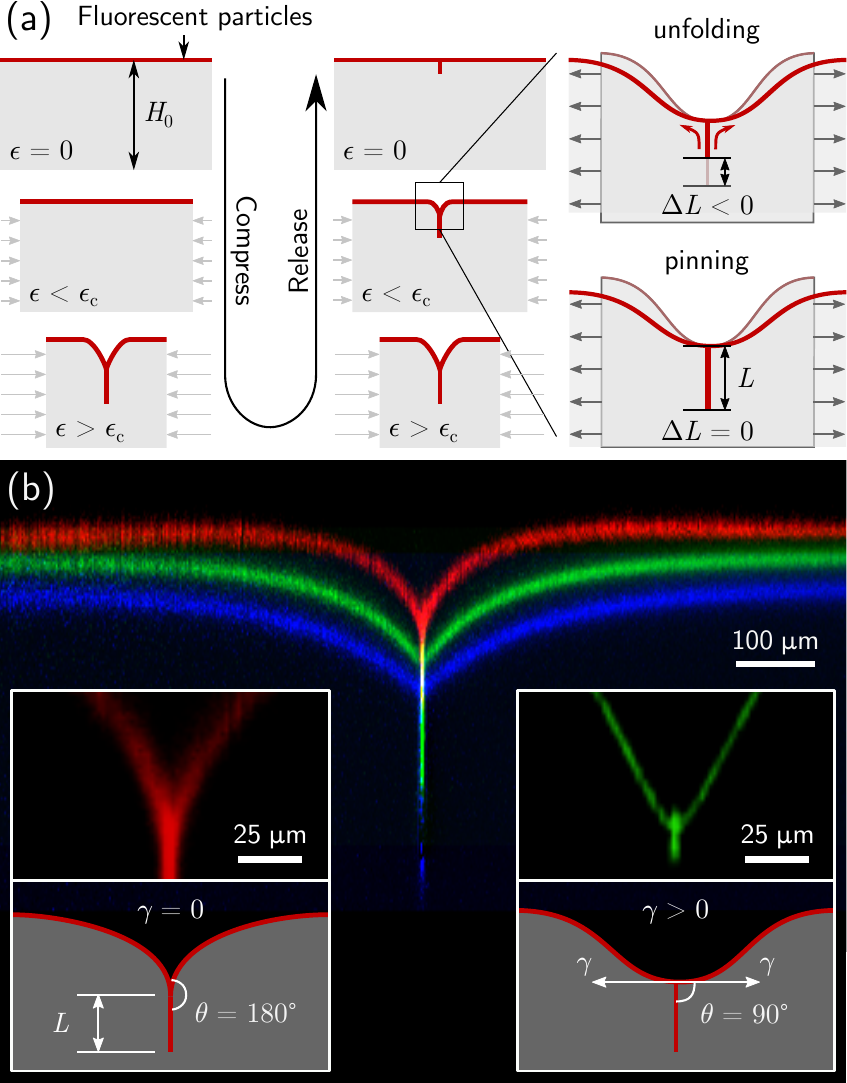}%
	\caption{(a) A uniaxially compressed soft material creases beyond a certain critical strain $\epsilon_c$, folding its surface into a self-contact. The crease persists below $\epsilon_c$, and ``scars" remain even at $\epsilon=0$: contact line pinning prevents complete unfolding. (b) Confocal images of creases for different elastocapillary lengths, $\ell = \gamma/G \approx 0$ (red), $\SI{20}{\micro\meter} $ (green), and $ \SI{300}{\micro\meter}$  (blue). Insets: solid surface tension impacts the contact angle $\theta$ at the edge of the self-contact.
	}%
	\label{fig:setup}%
\end{figure}%

The canonical creasing behavior can be realized by uniaxially compressing a slab of a soft material (\autoref{fig:setup}~a). Sharp creases form via a sub-critical bifurcation after a reaching a critical strain $\epsilon_c$~\cite{gent1999rubber,hong2009formation}.
When subsequently releasing the strain to below $\epsilon_c$, the length of the crease is only gradually reduced. This implies bistability since either creased or homogeneous states can be  found, depending on the deformation history~\cite{yoon2010swelling,chen2012surface,karpitschka2017}. Recent studies clarified the onset of creasing~\cite{hong2009formation,ciarletta2018matched,ciarletta2019nucleation}, invoking surface tension~\cite{yoon2010swelling,chen2012surface,liu2019elastocapillary} or the presence of a skin~\cite{hohlfeld2011unfolding} to explain the observed bistability. 

Interestingly, a microscopic residual crease typically remains even after the strain is fully released to $\epsilon=0$ (\autoref{fig:setup}~a). This small feature, referred to as a ``scar",  is of great significance, since it serves as a nucleus for creases when repeating the compression~\cite{trujillo2008hydrogels,yoon2010swelling,chen2012surface}. As such, these scars endow soft materials with mechanical memory~\cite{trujillo2008hydrogels,yoon2010swelling}, offering a potential of dynamic programmability of surface folds. 
Despite their importance, scars have remained somewhat enigmatic. It was found that scars are not due to material failure, and their persistence was argued to originate from adhesion~\cite{yoon2010swelling,chen2012surface,cai2012elastomer}. 
However, it is not clear whether adhesion and surface tension can actually lead to a reduction of surface energy compared to the flat, scarless state.
Recent work focused on the consequences of surface tension to the onset bifurcation of creasing~\cite{liu2019elastocapillary}, but it is not known how surface tension affects folding and unfolding at the microscale.

In this Letter we resolve the micro- and macro-morphology of adhesive creases by confocal microscopy (\autoref{fig:setup}~b), and identify the role of surface tension ($\gamma$), inside the self-contact. It is found that surface tension induces a second fold at the edge of the self contact, turning the surface into a T-shaped profile (\autoref{fig:setup}~b, bottomright). This involves a change of the contact angle $\theta$ from $\SI{180}{\degree}$ for $\gamma\sim 0$ to $\SI{90}{\degree}$ for $\gamma > 0$.
Further, we show that folding and unfolding the crease are, at the microscale, intrinsically asymmetric processes. The unfolding is inhibited, and ultimately prevented, by contact line pinning. This pinning-induced hysteresis implies a new type of bistability, even far above the onset of creasing, and offers a natural explanation for the formation of scars.

\paragraph{Experimental}%
A layer of a soft polymer gel (Dow Corning CY52-276, components A:B mixed 1:1 or 1.4:1 to obtain different shear moduli, thickness $H_0 \sim 1.0$ to $\SI{1.3}{\milli\meter}$) was prepared on top of a stiff, uniaxially pre-stretched PVS rubber sheet (Zhermack Elite Double 20).
To compress the gel layer, the pre-stretch of the support was slowly released by a micrometer~\cite{mora2011instability,cai2012elastomer}.
Fluorescent particles (Invitrogen FluoSpheres, $\SI{100}{\nano\meter}$ diameter) were added to the bottom and top surface of the gel.
Gel thickness and top surface morphology were measured with an upright confocal microscope (Leica TCS-SP2) with 10X and 40X magnification.
To image the self-contact through the free surface with minimal optical artifacts, we used index-matched immersion liquids.
The immersion also offers a way to tune the gel surface tension: We measured $\gamma\sim 0$ for the immersion with a long-chain silicone oil (Wacker, \SI{12.5}{\pascal\second}) and $\gamma\sim\SI{20}{\milli \newton \per \meter}$ for a water-glycerol mixture~\cite{cai2012elastomer}.
The relative importance of surface tension $\gamma$ to shear modulus $G$ is quantified by the elastocapillary length, $\ell = \gamma/G$. Importantly, these liquids caused virtually no swelling of the gel, as was tested by prolonged immersion. The typical compression protocol of an experiment is sketched in \autoref{fig:setup}~a. First we increased the strain $\epsilon$ in small steps, recording after each step and a prolonged waiting time ($\gtrsim \SI{15}{min}$, much larger than the material relaxation time $\lesssim\SI{0.5}{\second}$) the free-surface morphology by an xyz-scan.
After creases had formed we compressed a bit further, then repeated this procedure while decreasing $\epsilon$ again.
Further experimental details can be found in the supplemental material~\cite{suplmat}.

\paragraph{Elastocapillary self-contact}%
We first quantify the morphology of the elastocapilary self-contact, and how it is altered by surface tension. \autoref{fig:setup}~b shows creased surface profiles from three experiments with different elastocapillary lengths: $\ell = \gamma/G \sim 0$ (red), $\sim\SI{20}{\micro\meter}$ (green, $G\sim\SI{1.1}{\kilo \pascal}$), and $\sim\SI{300}{\micro\meter}$ (blue, $G\sim\SI{65}{\pascal}$).
Increasing the elastocapillary length amplifies the relative importance of surface tension, which on the macroscale leads to shallower indentations. 

However, the most important consequence of $\gamma > 0$ is reflected in the micro-morphology of the contact region (high magnification data, insets of \autoref{fig:setup}~b). The angle $\theta$ that the surface describes at the contact line changes from a gentle touchdown with $\theta=\SI{180}{\degree}$ for $\gamma\sim 0$, to $\theta=\SI{90}{\degree}$ for $\gamma>0$. 
Hence, besides the fold at the bottom of the crease, there is a second fold a the top of the self-contact where the surface profile acquires a T-shape. 
On scales much smaller than $\ell$, capillarity will be dominant over elasticity and we expect the contact angle to be given by Neumann's law~\cite{Karpitschka:SM2016,Xu:NC2017,Pandey:PRX2020}.
In case of perfect self-adhesion, which is expected for soft polymer gels, the surface tension in the self-contact vanishes completely.
The remaining balance of gel-vapor surface tensions implies the fold of $\theta = \SI{90}{\degree}$ (\autoref{fig:setup}~b), in good agreement with our experiments. 
 	
\begin{figure}[tb!]%
	\centering\includegraphics{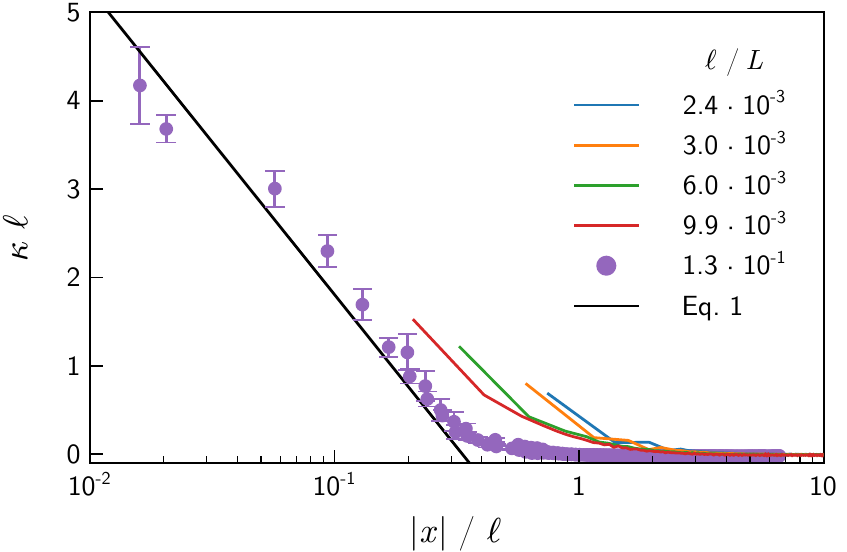}%
	\caption{Free surface curvature $\kappa$ versus the horizontal distance $x$ to the contact line, in units of the elastocapillary length $\ell=\gamma/G$. $\kappa$ diverges logarithmically for $x\rightarrow 0$, both in experiments (symbols) and in simulations (lines). The horizontal shift between the curves reflects difference in gauge pressure $p_0$ in the $\log$, which depends on the ratio $\ell/L$. The experimental data was taken after compressing, for $\ell\sim\SI{20}{\micro \meter}$.}%
	\label{fig:curv}%
\end{figure}%

Consequential to this second fold is a strong curvature $\kappa$ of the free surface near the contact line. When attempting to quantify this curvature, however, it turns out that $\kappa$ does not reach a well-defined limiting value. Rather, it still grows as we reach our measurement resolution, which is well below the elastocapillary length. \autoref{fig:curv} (purple markers) shows the measured curvature as a function of the distance to the contact line, for $\ell\sim\SI{20}{\micro\meter}$. Surprisingly, the data suggest a logarithmic divergence of $\kappa$ as the contact line is approached. 

The logarithmic singularity of curvature is caused by the capillary boundary condition, which forces the material into a $\SI{90}{\degree}$ angle. Analogous to the bottom of the crease~\cite{hong2009formation,karpitschka2017,ciarletta2018matched}, a fold of angle $\theta$ introduces a weak (logarithmic) stress singularity. For a neo-Hookean solid the stress singularity reads $p_{\rm el}= (\pi/\theta - \theta/\pi)\, G \ln(|x|) + p_0$~\cite{singh1965note}, where $\theta$ is the fold angle in radians and $p_0$ a gauge pressure. At the bottom of the self contact, where the crease was initiated, the fold has an angle $\theta=2\pi$. In contrast, the two folds at the contact line involve an angle $\theta=\pi/2$, connecting the self-contact to the free surface in a T-shape. Thus the elastic stress must be balanced by the Laplace pressure $\gamma \kappa = -p_{\rm el}$, which for a right angle gives the elastocapillary balance
\begin{equation}
	\label{eq:inner}
	\gamma \kappa \simeq \frac{3}{2} G \ln{\left(b\frac{\ell}{|x|}\right)},
\end{equation}
valid at distances $|x| \ll \ell$. Here we absorbed the gauge pressure $p_0$ into a dimensionless constant $b$. The prediction (\ref{eq:inner}) is shown as the solid line in \autoref{fig:curv}, in excellent agreement with the experimental data. We only adjusted the value of $b$, which cannot be derived by the local analysis of the singularity: it reflects a gauge pressure that is inherited from the full solution, invoking scales larger than $\ell$.

To explore this effect, we performed finite element simulations of elastocapillary creases, assuming a neo-Hookean solid with constant surface tension (see supplemental material for details~\cite{suplmat}). Numerical results are obtained for various ratios of $\ell/L$, where $L$ is the crease length. All numerical results exhibit the logarithmic divergence of curvature (\autoref{fig:curv}, lines), following the prediction (\ref{eq:inner}). The fact that the data are shifted laterally reflects the non-universality of $b$. Still, the numerical data gradually approach the experimental data as $\ell/L$ approaches the experimental value. 

Thus we conclude that the capillary nature of adhesive creases has two important consequences at the microscale: (i) it governs the contact angle $\theta$, implying a second fold at the top of the crease where the surface is T-shaped. (ii) the fold, in turn, introduces a logarithmic singularity of the elastic stress and the surface curvature at the edge of the self-contact.

\begin{figure}%
	\centering\includegraphics{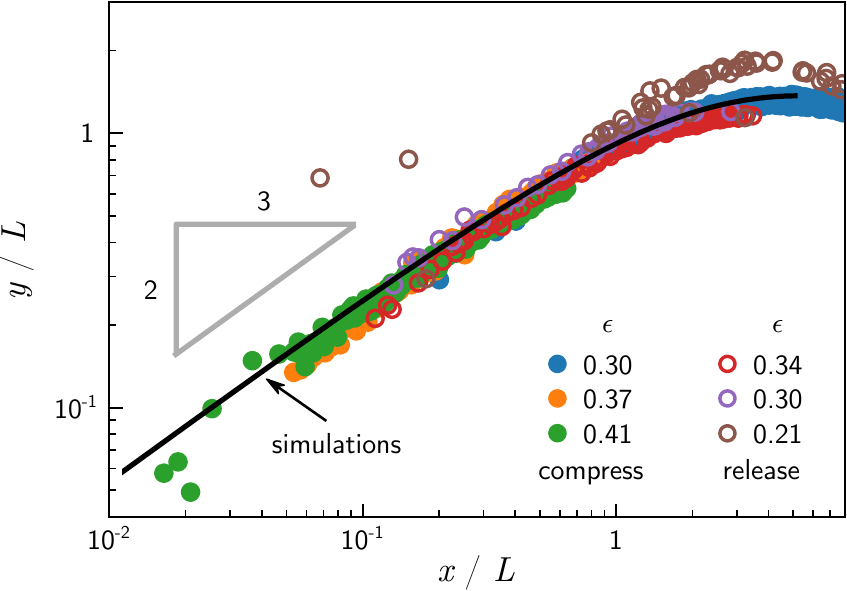}%
	\caption{Free surface profiles (experiment and simulation) for various $\epsilon$ for the case where $\gamma \approx 0$. The collapse of data confirms a universal shape that is governed by the crease length $L$, according to (\ref{eq:intermediate}).}%
	\label{fig:23power}%
\end{figure}%

\paragraph{Intermediate asymptote}%
Contrarily, for  $x\gg \ell$, we expect capillary effects to play no role and hence, the problem should be described by a purely elastic solid mechanics. Then the crease length $L$ becomes the relevant scale, and the morphology of the free surface was predicted to exhibit the scaling~\cite{karpitschka2017}

\begin{equation}
	\label{eq:intermediate}
	\frac{y-y_0}{L}\sim \left(\frac{x}{L}\right)^{2/3},
\end{equation}
where $y_0$ is the vertical coordinate of the contact line. This intermediate asymptote is expected to be valid whenever $\ell \ll x \lesssim L$. 
\autoref{fig:23power} shows the collapse of simulations and measurements for $\gamma\sim 0$ and various $\epsilon$, and confirms the 2/3 exponent.

\paragraph{Folding-unfolding asymmetry}%
Elastocapillary (inner) and elastic (intermediate) regions can be collapsed simultaneously by choosing the appropriate  scales $x = {l}_x \hat{x}$ and $y = {l}_y \hat{y}$ on each axis.
The inner, elastocapillary morphology (\ref{eq:inner}), requires ${l}_x^2/{l}_y = \ell$ to collapse the data. The scaling behavior of the intermediate region (\ref{eq:intermediate}), requires ${l}_y/{l}_x^{2/3} = L^{1/3}$.
Both requirements are fulfilled simultaneously by choosing

\begin{equation}\label{eq:rescaling}
{l}_x=\ell^{3/4}L^{1/4},\quad{l}_y=\ell^{1/2}L^{1/2}.
\end{equation}

\begin{figure}%
	\centering\includegraphics{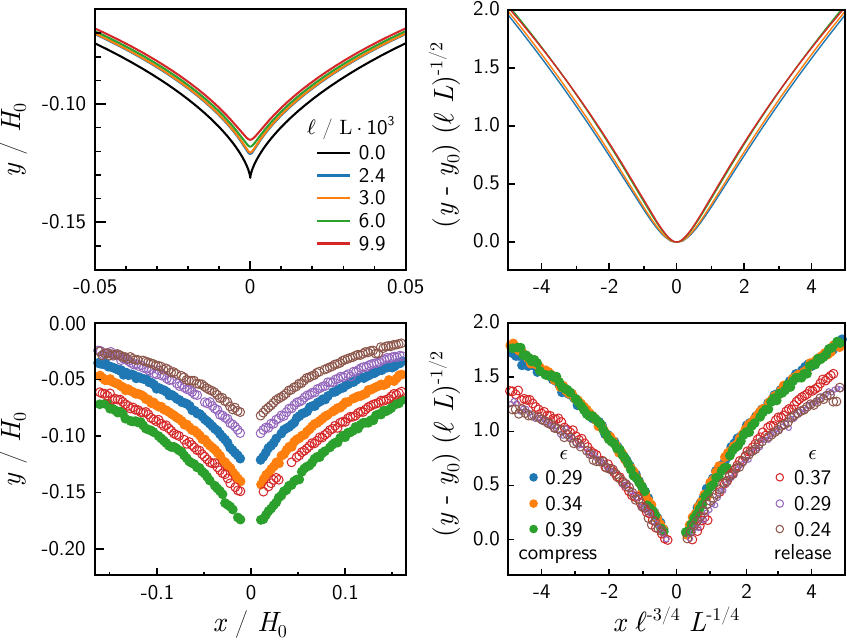}%
	\caption{Folding-unfolding asymmetry. Numerical (top) and experimental (bottom) profiles of the crease can be collapsed after rescaling by (\ref{eq:rescaling}) (right panels). The experimental curves (elastocapillary length $\ell \sim\SI{20}{\micro\meter}$) cluster into two sets, depending on whether the global compressive strain $\epsilon$ was increased (filled symbols) or decreased (open symbols) prior to the measurement. Surface profiles are shallower during unfolding than during folding.}
	\label{fig:rescaled}
\end{figure} 

In \autoref{fig:rescaled}, the original interface profiles (left panels) are shown next to the profiles as rescaled by the prediction (\ref{eq:rescaling}) (right panels). The simulation data (top) almost collapse on a single curve, where the remaining difference can be attributed to the non-universal gauge pressure.
The experimental data (bottom), by contrast, clusters into two groups. Filled symbols correspond to measurements after $\epsilon$ has been increased and form the upper master curve (folding). Open symbols correspond to measurements after $\epsilon$ has been decreased and form the lower master curve (unfolding). 

The experiments show that the history of the crease is important for the observed morphology, even for $\epsilon> \epsilon_c$ where only the creased state is stable. The interface profiles observed during folding are manifestly different from the profiles during unfolding, the latter being more shallow. This asymmetry between folding and unfolding is not observed in the simulations. For a given set of parameters, the numerical minimisation of elastocapillary energy selects a unique crease morphology. Clearly, an element beyond equilibrium elastocapillary mechanics is required to properly interpret the experiments, which originates from contact line physics. 

\begin{figure}[tb!]%
	\centering\includegraphics[width=\columnwidth]{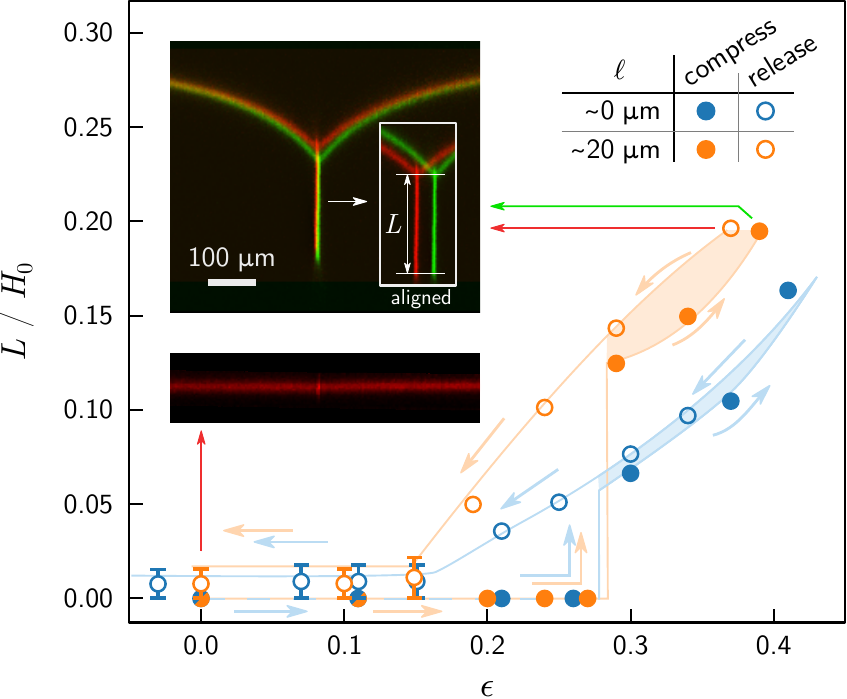}%
	\caption{Length of the self contact $L$ as a function of global strain $\epsilon$, for purely elastic ($\ell\sim 0$, blue) and adhesive, elastocapillary ($\ell\sim\SI{20}{\micro\meter}$) creases. Two types of hysteresis can be observed: (i) the subcritical/critical transitions from the homogeneous to the creased state ($\epsilon\sim 0.28$), and back to homogeneous ($\epsilon\sim0.15$), respectively. (ii) The creased state exhibits a manifold of possible $L$ for each $\epsilon$ (shaded area). $L$ depends on its history, reminiscent of contact line pinning. Upper inset: Two profiles of similar $L$ but different $\epsilon$ (green: after a compression; red: after releasing). Lower inset: A tiny scar remains after full relaxation and even at $\epsilon<0$.
	The error bars in the main plot reflect the uncertainty in determining the absolute size of the (residual) crease. As this merely provides a systematic offset to the length of developed creases, the uncertainty of the length hysteresis (shaded area) is much smaller ($\sim 0.005$).
	}%
	\label{fig:bifurcation}%
\end{figure}%

\paragraph{Contact line pinning}%
The difference in morphologies upon switching from compression to expansion is visualized in the inset of \autoref{fig:bifurcation}. The green profile is taken during the compression phase, while the red profile is obtained after a subsequent release of strain. Clearly, the global indentation of the free surface decreased upon reducing $\epsilon$, while the crease length $L$ remained identical up to the measurement resolution.
This offers direct evidence for contact line pinning, where there is no ``unfolding" at all, i.e. no change of material points at the contact line (cf. sketch in \autoref{fig:setup}~a). 

In the present context, the morphological hysteresis readily impacts the crease length $L$, which is no longer a pure function of the imposed strain $\epsilon$. This is shown in \autoref{fig:bifurcation} (main plot), reporting $L$ versus $\epsilon$. While bistability is well-known below the onset of creasing, we here find an additional bistability that is caused by contact line pinning: the shaded area shows that, even well above the onset of creasing, the crease length still exhibits a history dependence. After increasing $\epsilon$, smaller $L$ are selected, rather than after decreasing $\epsilon$, especially for $\ell \sim\SI{20}{\micro\meter}$ (orange curve, same data as in \autoref{fig:rescaled}).
For $\gamma\sim 0$, the measurement error is comparable to the detected hysteresis, and no definitive statement on its existence can be made. Due to experimental limitations we can only give an upper bound for the scar length $\sim\SI{10}{\micro\meter}$. However, the fact that this hysteresis loop becomes more pronounced for larger $\ell$ suggests a capillary origin of contact line pinning. 

The observed contact line pinning is analogous to the wetting of liquids on solid surfaces~\cite{BonnRevModPhys2009}. In that case, the motion of solid/liquid/vapor three-phase contact lines is arrested by pinning on heterogeneities of the solid surface. Such heterogeneities lead to a complicated energy landscape which allows for a range of stable liquid morphologies, leading to a range of contact angles. The pinning observed for creasing could be of similar origin as in wetting, where an energetic barrier $\Delta\gamma$ is required to move the contact line across features of the surface topography. 
We therefore expect a range of mechanically stable crease lengths of the order of $\Delta L \sim \Delta \gamma/G$. This ultimately prevents complete unfolding: the remaining elastic energy is not sufficient to overcome this pinning-barrier. This explanation for scars is similar to that given in \cite{chen2012surface}, which was phrased in terms of an energy release rate rather than $\Delta \gamma$. From our findings, however, is clear that the energy release rate for unfolding an adhesive crease cannot be given by the reversible work of self-adhesion, based on surface energies, as that would not offer a mechanism for contact line pinning. Indeed, our finite element simulations (with reversible adhesion) do not give a folding-unfolding asymmetry. They lack the new bistability indicated by the shaded areas in \autoref{fig:bifurcation}, and do not admit any scars. To exclude the tracer particles as cause for the scars in our experiments, we also tested an uncoated specimen by brightfield reflection microscopy, finding the same behavior (see supplemental material~\cite{suplmat}). Our explanation in terms of contact line pinning is also consistent with the observation of scar-annealing on long timescales~\cite{cai2012elastomer,chen2012surface}: once again, this resembles the case of liquid wetting, where contact line pinning can indeed be overcome by thermal activation \cite{Rolley:PRL2007,Snoeijer:ARFM2013,Perrin:PRL2016}.

\paragraph{Outlook}%
Our confocal microscopy experiments have revealed an intrinsic folding-unfolding asymmetry, induced by contact line pinning, that offers a natural explanation for scars. Pinning is therefore a central element in creasing that needs to be accounted for in theory and simulations, and offers a way to articulate the role of ``defects". In addition, we have shown that surface tension dictates the mechanics at scales below the elastocapillary length, folding the solid into well-defined contact angles. As such, the crease morphology opens a new route to quantify solid-solid interfacial mechanics, in line with recent developments for solid-liquid interfaces \cite{Xu:NC2017,Andreotti:ARFM2020}.
Understanding the interfacial micro-mechanics of soft materials unfolds their potential as programmable matter~\cite{Hawkes:PNAS2010}, impacting for instance soft robotics~\cite{Rus:N2015}, biomolecular patterning~\cite{kim2009patterns}, or smart textiles~\cite{Persson:AM2018}

\begin{acknowledgments}
S.K. acknowledges funding from the German research foundation (DFG, project no. KA4747/2-1), and M.H.E. and J.H.S. from NWO through VICI Grant No. 680-47-632.
\end{acknowledgments}

\newpage
\onecolumngrid
\appendix
\newpage

\section*{Supplemental Material}

\renewcommand{\thefigure}{S\arabic{figure}}
\setcounter{figure}{0}

\section{Experimental}
The stiff support layer was made of Zermack Elite Double 20 in a 1:1 ratio, resulting in a shear modulus $G$ of \SI{0.1}{\mega \pascal}, which was two orders larger than the stiffness of the soft layer. The layer was cured in a oven at \SI{80}{\celsius} after degassing in vacuum. Then, the support layer was coated with fluorescent particles (Invitrogen FluoSpheres, orange fluorescent, \SI{100}{nm} diameter) at low coverage, to label the support-gel interface. Prior to applying the the soft layer, the support layer was clamped in a movable stage and stretched by \SI{100}{\percent}.

The soft layer was made of a polydimethylsiloxane gel (Dow Corning CY52-276, components A and B) and was poured after degassing in vacuum onto the pre-stretched stiff layer. The sample was then heat-cured in an oven at \SI{75}{\celsius} for three hours. Two types of soft layers were prepared: a 1:1 mixing ratio of components A and B resulted in $G\sim\SI {1.2}{\kilo \pascal}$. A mixing ratio of 1.4:1 (A:B) resulted in $G\sim\SI {65}{\pascal}$. All moduli were measured with an Anton Paar MCR502 Rheometer and a $\SI{50}{\milli\meter}$ parallel plate geometry. After curing, a layer of fluorescent tracer particles (Invitrogen FluoSpheres, orange fluorescent, \SI{100}{nm} diameter) was deposited onto the soft layer.

Two immersion liquids were used during the experiments: Wacker $\SI{12.5}{\pascal\second}$ silicon oil was used for the $\gamma\sim\SI{0}{\milli\newton\per\meter}$ measurements, whereas an index-matched water/glycerol mixture to obtain $\gamma\sim\SI{20}{\milli\newton\per\meter}$. Prior to measurement, the liquid was put onto the soft layer and covered by a large  microscope cover slide to prevent evaporation and observation through a curved meniscus. The slide rested on small glass beads of 2 mm diameter to keep it separated from the sample.

An upright confocal  microscope (Leica TCS-SP2) was used for imaging, operating with either a 10X or a 40X objective. Scanning volumes were $(3.0\times 3.0\times 3.4)\;\SI{}{\cubic\micro\meter}$ (10X objective)  and $(0.7\times 0.7\times 0.7)\; \SI{}{\cubic\micro\meter}$ (40X objective), where the last number refers to the vertical axis.  

\begin{figure}[tb]
	\centering\includegraphics[width=.9\textwidth]{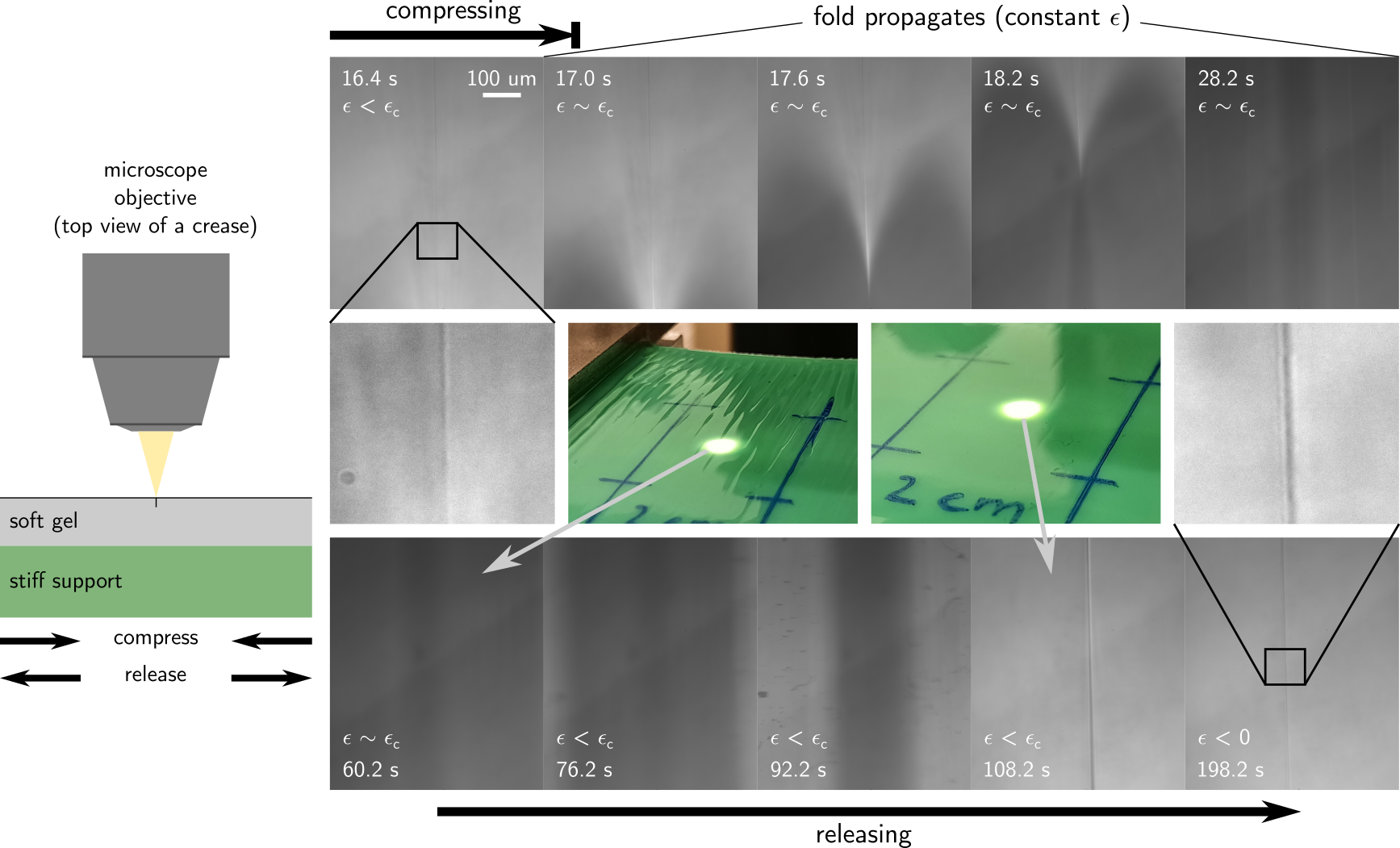}%
	\caption{\label{fig:supp1}Top view images of a scar that creases (top row) and the unfolding of the same crease down to a scar (bottom row), taken by brightfield reflection microscopy during the second compression of a sample without tracer particles at the gel surface. Contrast is generated by refraction at the gel surface: inclined surface regions are imaged darker than horizontal ones. Top row: After reaching a critical compression, the crease folds spontaneously along a scar, which was left behind by a crease from the previous compression. The crease folds in a zipping motion, visualized by the propagating contrast.  Bottom row: Unfolding upon releasing the compression is quasi-stationary and homogeneous along the fold. The last image was recorded at slightly negative compression, showing that the scar persists.}
\end{figure}

We performed another experiment on a similar specimen, but without the tracer particles, using brightfield reflection microscopy, where the (residual) creases induce contrast by refraction at the gel surface. Figure~\ref{fig:supp1} shows micrographs of scars and creases during the second compression, together with macro images of the specimen.

\section{Numerical methods}
The numerical simulations of the adhesive crease are performed using the Finite Elements Method (FEM) library \texttt{oomph-lib}\cite{oomph}. These simulations are based on those in \cite{karpitschka2017}, with the addition of surface tension. A square slab of material of Lagrangian height $H$ and width $W=H$ is compressed in horizontal directions. The material is simulated using a refineable mesh of Neo-Hookean solid elements, refined according to a fixed refinement pattern. Around the self-contact, the the mesh resolution was finest ($\Delta x = 9.8\times 10^{-4}L$) and increased toward the domain boundaries to $\Delta x = 0.25L$. Non-dimensionalisation using the crease length allows us to probe the full range of solutions by variation of $W$ and $\gamma$ only.

The bottom surface is constrained in the vertical direction and has a no-shear condition in the horizontal direction. The left edge of the material is constrained to the symmetry axis in horizontal direction, with a no strain (symmetry) condition in the vertical direction. Similarly, the right edge has a no-shear condition in vertical direction, but is constrained to a variable position horizontal direction to impose the global compression through Lagrange multipliers. The top surface is split into two parts: $X\leq1$ will be part of the self-contact, and $X\geq1$ will form the free surface. For details, we refer to~\cite{karpitschka2017}. On the free surface the Laplace pressure $\gamma$ is applied, with a contact angle of 90 degrees at either side. The self contact is subject to a no-shear condition in vertical direction, and is constrained to the symmetry axis by Lagrange multipliers. 

A small cutout is made at the interface between the left boundary and the self contact, as detailed in \cite{karpitschka2017}, to avoid the stress singularity at the sharp fold \cite{singh1965note}. Such a cutout cannot be used at the interface between the self-contact and the free surface, as the morphology of the free surface at this point is of particular interest. Therefore, the elements neighboring the singularity are formulated in polar coordinates around the contact line. 
In the vertical direction, the residual at the contact line is only controlled by surface tension and an imposed force that reflects any residual bulk elastic stresses.
The latter force is an unknown of the problem, and determined in order to fulfill a bulk element with a \SI{45}{\degree} angle at the contact line, ensuring a symmetry necessary to find a consistent solution \cite{Pandey:PRX2020}. This force scales approximately as $\Delta x \log(\Delta x)$ for various mesh sizes $\Delta x$ and should hence be seen as a consequence of the finite resolution of a numerical simulation. Finally, the location of the right boundary $w$ is controlled by requiring a \SI{90}{\degree} contact angle.

\end{document}